\documentclass[a4paper,10pt,twoside]{cpc-hepnp}

\usepackage{multicol}
\usepackage{graphicx}
\usepackage{booktabs}
\usepackage{amssymb,bm,mathrsfs,bbm,amscd}
\usepackage[tbtags]{amsmath}
\usepackage{lastpage}
\usepackage{CJK}

\begin{document}
%\begin{CJK*}{GBK}{song}

\fancyhead[c]{\small Chinese Physics Letter~~~Vol.1, No. 1 (2019)
111111} \fancyfoot[C]{\small 010201-\thepage}

%\footnotetext[0]{Received 14 March 2009}

\title{Topological Field Theory and Phase Transition\thanks{Supported in part by the
National Natural Science Foundation of China (11775118 and 11535005). }}

\author{
%\quad ZHOU Jing(Öܾ©)$^{2;2)}$
\quad Jing Zhou
\quad Jia-Lun Ping$^{1}$\email{jlping@njnu.edu.cn}
}
\maketitle

\address{%
%$^1$ Institution or University where the author works,  district,  postal code,  country\\
%$^2$ {\bf Example}: Institute of High Energy Physics, Chinese Academy of Sciences, Beijing 100049, China\\
Department of Physics, Nanjing Normal University, Nanjing, Jiangsu 210097, China
}

%\begin{abstract}
\emph{ The partition function of the topological twisted super Yang-Mills field theory on the boundary can be expanded as Jones polynomial, which can be computed as expectation values of Wilson loop operators. We show that the zero of the Jones polynomial is Lee-Yang type. Moreover, Lee-Yang phase transition is also discussed in the Jones polynomial of torus
knot and the topological twisted super Yang-Mills field theory.}
%\end{abstract}

%\begin{keyword}
%topological twisted super Yang-Mills field theory£¬Lee-Yang£¬Jones polynomial
%\end{keyword}

%\begin{pacs}
%1--3 PACS(Physics and Astronomy Classification Scheme, http://www.aip.org/pacs/pacs.html/)
%\end{pacs}

\footnotetext[0]{\hspace*{-3mm}\raisebox{0.3ex}{$\scriptstyle\copyright$}2019
Chinese Physical Society and IOP Publishing Ltd}%

\begin{multicols}{2}

%\section{Introduction}
The Jones polynomial~\cite{Jones:1985dw,Jones:1987dy} is a celebrated invariant of a knot in the
three-dimensional space, it is discovered by Jones as an shoot of his work on von Neumann algebras.
Then many descriptions and generalizations of the Jones polynomial were discovered immediately after
Jones's work. In fact, the polynomial has multiple relations to many aspects of mathematical physics,
which including statistical mechanics, two dimensional conformal field theory, representations of
braid groups and three-dimensional Chern-Simons gauge
theory~\cite{Freyd:1985dx,Turaev:1988eb,Jones:1989ed,Akutsu:1987bc,Witten:1988hf}.

One of the most remarkable known quantum field theories in four dimensions is the N = 4 supersymmetric Yang-Mills theory. This theory has the largest supersymmetry for a four-dimensional theory without gravity. A long-standing conjecture asserts that this theory has a symmetry exchanging strong and weak coupling and exchanging electric and magnetic fields. It was realized by Vafa and Witten ~\cite{Vafa:1994tf}through topological twisted the super Yang-Mills theory. Khovanov homology [9] is considered to be a topological theory in four dimensions super Yang-Mills field theory. The relation between the Khovanov homology and the knot is that the four-dimensional theory associated to Khovanov homology, when compactified on a circle, reduces to the three-dimensional theory that yields the Jones polynomial.

In 1952, Lee-Yang~\cite{Yang:1952be,Lee:1952ig} established a rigorous relation between the analytic
properties of free energies and thermodynamics through continuation of the partition function to the
complex plane of physical parameters. And they considered a general Ising model with the ferromagnetic
interaction $J_{ij}>0$ under a magnetic field $h$ with the Hamiltonian. Then they proved that all the
zeros of this partition function lie on the unit circle in the complex plane of $z$. And one can find
that the Lee-Yang zeros characterize the analytic properties of partition function and the thermodynamics
systems. So, determining the Lee-Yang zeros is not only useful for a complete picture of thermodynamics
and statistical physics but also important for studying phase transition of physics system. Surprisingly,
Lee-Yang zeros are observed by measuring quantum coherence of a probe spin coupled to an Ising-type spin
bath~\cite{xinhua:2015hf}. In other word, the zeros really exist.

Actually, Lee-Yang zero has go beyond the statistical physics. Maloney and Witten~\cite{Maloney:2007ud} shown
that the Hawking-Page transition~\cite{Hawking:1982dh} in $AdS_{3}$ space is Lee-Yang type, while the original
Lee-Yang phase transition is only for two dimensional Ising model. This is an important reason that we argue
that the phase transition in topological field theory is Lee-Yang type. Next, let us interpret the Lee-Yang
phase transition in the $AdS_{3}$ space in more detail.

The Hawking-Page transition can be seen from Lee-Yang condensation of zeros in the partition function for
$k\rightarrow\infty$. Actually, the partition function $Z(\tau)$ of three dimensional gravity is a modular
function which computes at fixed temperature Im$\tau$ and angular potential Re$\tau$. In the limit of infinite
volume, these zeroes condense along the phase boundaries. So, it gives rise to phase transition. The analog of
the infinite volume limit for the partition function $Z(\tau)$ is $k\rightarrow\infty$ because as we know
$k=\ell / 16 G$ implies that k directly proportional to the $AdS_{3}$ radius. In this case, the partition
function $Z(\tau)$ is non analytic which corresponding to the occurrence of phase transition.

Our aim here, however, is not to understand the phase transition in $AdS_{3}$ space but the zeros in Jones polynomial
or topological field theory. Then, there is a natural question to ask: What do these zeros mean? Or, in other word,
can phase transition happen in topological field theory? In the following work, we show that it may shed light on this issue.

%\section{Title and author}
%Funding information is put behind the title, use $ \backslash
%$thanks\(\) command. When making an entry of the author's name,
%attention should be paid to the emblem of the author's institution.
%Different to the command in revtex 4, ``Chinese Physics C" uses
%$\backslash $danwei\(\) command in writing the author's institution.
%To avoid a mistake at this point, more attention should be called
%for.

%\section{Lee-Yang type phase transition of super Yang-Mills theory }
The Chern-Simons action for a gauge theory with gauge group $G$ and gauge field $A$ on an oriented three-manifold
$M$ can be write as ~\cite{Witten:2011zz},
\begin{equation}
 CS(A)= \frac{k}{4\pi}\int_{M}Tr\left(dA\wedge A+\frac{2}{3}A\wedge A\wedge A\right).
\end{equation}
Here $k$ is an integer for topological reasons. All we really need to know for now about $CS(A)$ is that
it is gauge-invariant. The Feynman path integral now is formally an integral over the infinite-dimensional
space of connections $A$. This is a basic construction in the quantum field theory,
\begin{equation}
\int DA\exp\left(iCS\right).
\end{equation}
To including a knot which it is an embedded oriented loop $K \subset M$, we make use of the loop of the
connection $A$ around $K$. Picking an irreducible representation $R$ of $K$, then one can associate an
observable, the trace of Wilson loop operator,
\begin{equation}
 W\left(K,R\right)= Tr P\exp\int_{K} A.
\end{equation}
Then we define a natural invariant of the pair ($M$, $K$), So the partition function can write as,
\begin{equation}
 Z\left(M; K\right)= \int DA\exp\left(iCS\right)\prod_{i=1}^{r} W\left(K_{i},R\right).
\end{equation}
This is a topological invariant of the knot $K$ in the three manifold $M$, which depends only on $G$, $R$ and $K$.
Then, Witten made use of the topological invariance of the theory to solve Chern-Simons topological theory on
three manifold $M$ with collection of knots, which is a suitably normalized Chern-Simons partition function,
\begin{equation}
  \langle\mathcal{W}(K, R)\rangle= \frac{\int DA\exp\left(iCS\right)\prod_{i=1}^{r}
  W\left(K_{i}, R\right)}{\int DA\exp\left(iCS\right)},
\end{equation}
equals to the Jones polynomial,
\begin{equation}
\left\langle\mathcal{W}(K, R)\right\rangle= J_{K}(q).
\end{equation}
For modular transformation and $G=S U(2)$ we have,
\begin{equation}
S_{m n}=\sqrt{\frac{2}{k+2}} \sin \left(\frac{(m+1)(n+1) \pi}{k+2}\right).
\end{equation}
%By surgery operation $S$ we obtain the partition function on
And we obtain the partition function on
$S^{3}$~\cite{Witten:1988hf,Gopakumar:1998ki},
%\begin{equation}
%Z\left(S^{3}\right)=\sum_{j} S_{0}^{j} Z\left(S^{2} \times S^{1} ; %R_{j}\right)=S_{0,0}=\sqrt{\frac{2}{k+2}} \sin \left(\frac{\pi}{k+2}\right)
%\end{equation}
\begin{equation}
Z\left(S^{3}\right)=S_{00}=\sqrt{\frac{2}{k+2}} \sin \left(\frac{\pi}{k+2}\right).
\end{equation}
%The general formula for $G=S U(N)$ is given by,
%\begin{equation}
%\begin{aligned}
%Z_{S^{3}}(S U(N), k)=&e^{i \pi N(N-1) / 8} \frac{1}{(N+k)^{N / 2}} \sqrt{\frac{N+k}{N}}\\ &\prod_{j=1}^{N-1}\left(2 \sin \frac{j \pi}{N+k}\right)^{N-j}.
%\end{aligned}
%\end{equation}

%\section{The Jones polynomial and topological twisted tuper Yang-Mills theory}
we have shown that the three-dimensional quantum theory gives a definition of
the Jones polynomial of a knot. An important fact is that we can expand the partition function to a
polynomial. Then it is natural to study the zeros of this polynomial. However this is not the whole
story, we need to know the Lee-Yang zero first. Let $Z$ be the canonical partition function which is
written as,
\begin{equation}
 Z=\sum_{i=1}^{n}e^{\frac{\mu N-E_{n}}{kT}},
\end{equation}
where $E_{n}$ is the energy, $\mu$ is the chemical potential and $N$ is the number of particles. If we expand
partition function $Z$ to a polynomial of $z$~\cite{blythe:12k},
\begin{equation}
 Z= 1+a_{1}z+a_{2}z^{2}+\cdots+a_{M}z^{M},
\end{equation}
where $z=e^{\frac{\mu}{kT}}$. Here one should note that $a_{j}$ $\left(j=1,2,\cdots,M\right)$ are all positive
real number. Now Let us consider the root of the function,
\begin{equation}
 Z\left(T,V,z\right)=0.
\end{equation}
Then from the above two equations, one can find that there is no positive real root for the equation.
If analytical extension to the complex plane, we find that all the zeros are on the unit circle which is
the celebrated Lee-Yang unit circle theorem. But it is not the whole story. In physics, phase transition
is a very interesting phenomenon. It is argued that phase transition can only happen when the zeros of the
partition function has real part. In this case, the partition function is not a analytic function now. This
comment is very important for the zeros of the of Jones polynomial.

Since both of the partition function of Lee-Yang and Jones polynomial can expand to polynomial. Then there
may be natural link between the two structures. Now Let us consider the zeros of the Jones polynomial.
From the previous section, one may find that the
expectation value of a Wilson loop in representation with spin $\frac{1}{2}$ is,
\begin{equation}
\langle\mathcal{W}(K,R)\rangle=\frac{Z_{M; K}}{Z_{S^{3}}}=\frac{S_{10}}{S_{00}}=q^{\frac{1}{2}}+q^{- \frac{1}{2}},
\end{equation}
where
\begin{equation}
 q=\exp\frac{2i\pi}{k+2}.
\end{equation}
If $\langle\mathcal{W}(K,R)\rangle=0$, then it is means,
\begin{equation}
 q^{\frac{1}{2}}+q^{- \frac{1}{2}}=0,
\end{equation}
So one can find that there is only one root $-1$ which is on the unit circle. The next case is the
expectation value of Hopf link with 2 spin $\frac{1}{2}$ Wilson loop which can write as
\begin{equation}
  \langle\mathcal{W}_{hopf}(K,R)\rangle= \frac{Z_{M, K}}{Z_{S^{3}}}=\frac{S_{11}}{S_{00}}=\left(q^{\frac{1}{2}}+q^{- \frac{1}{2}}\right)\left(q+q^{-1}\right).
\end{equation}
Then one can find there are three roots which are $-1$, $-i$, $i$. In fact, according to the Lee-Yang theorem,
then there is no phase transition. Actually, the expectation value of the Wilson loop equals to zero, then we
can rewrite the Eq.(5) as,
\begin{equation}
 \frac{\int DA\exp\left(iCS\right)\prod_{i=1}^{r} W\left(K_{i},R\right)}{\int DA\exp\left(iCS\right)}=0.
\end{equation}
Then what we are interested in is,
\begin{equation}
 \int DA\exp\left(iCS\right)\prod_{i=1}^{r} W\left(K_{i},R\right)=0.
\end{equation}
So, one can find the zero of the Jones polynomial is equal to the zero of
$Z\left(M;K\right)= \int DA\exp\left(iCS\right)\prod_{i=1}^{r} W\left(K_{i},R\right)$.
It equals to say that phase transition do not happen in the topological field theory in the previous two case.

The above has shown that there is no phase transition in the trivial knot. Then it is necessary to study some
nontrivial knots which phase transition could happen. Now Let us study the zeros of Jones polynomial of torus
knot $J(m,n)$~\cite{Labastida:2000zp} which distributes on $x$ axis. one should note that the simplest nontrivial
example of this knot is the $(2,3)$ torus knot which is also known as the trefoil knot. In physics, since
real-zeros of an equation usually represent observable values, it is interested to investigate them in advanced.
The Jones polynomial of torus knot write as,
\begin{equation}
J\left(m,n\right)=1-q^{m+1}-q^{n+1}+q^{m+n},
\end{equation}
If taking the limit of $n\rightarrow\infty$, this leads to $n$ for all $m$, then one can find
that all roots are distributed uniformly on the unit circle. In fact, for all Jones polynomials of torus $J(p,q)$,
there are two positive real-zeros with one is 1 and another is inside $1<r<2$\cite{Wu:2001}. Then by Lee-Yang theorem,
one can find that the two real zeros are phase transition points. In other word, phase transition happen in this
type topological field theory.

Topological field theory has been important for establishing symmetries and dualities in field
theory and string theory \cite{Witten:1988ze,Bershadsky:1993cx,Manschot:2017xcr}. A generalization of
electric-magnetic duality to Yang-Mills theory, known as $S$-duality, acts naturally on
Vafa-Witten theory \cite{Vafa:1994tf}. This duality, proposed by Montonen and Olive \cite{Montonen:1977sn},
states that Yang-Mills theory with gauge group $G$ and complexified coupling constant
\begin{equation}
\tau=\frac{\theta}{2 \pi}+\frac{4 \pi i}{g^{2}}
\end{equation}
has a dual description as Yang-Mills theory, whose gauge group is the Langlands dual group  and with inverse
coupling constant $-1 / \tau$ . Together with the periodicity of the $\theta$-angle, this generates the
$S L(2, \mathbb{Z})$ $S$-duality group,
\begin{equation}
%Z_{N}\left(\frac{a \tau+b}{c \tau+d}\right) \sim Z_{N}(\tau), \quad \left( \begin{array}{ll}{a} & {b} \\ {c} & {d}\end{array}\right) \in S L(2, \mathbb{Z})
Z\left(\frac{a \tau+b}{c \tau+d}\right) \sim Z(\tau), \quad \left( \begin{array}{ll}{a} & {b} \\ {c}
& {d}\end{array}\right) \in S L(2, \mathbb{Z}).
\end{equation}
%We will discuss later the richer structure of $SU(N)$ theories under the action of $SL(2, \mathbb{Z})$.
The Vafa-Witten twist of $N = 4$ super symmetric Yang-Mills theory with gauge group $SU(N)$ contains a commuting
BRST-like operator $Q$. For a suitable $I$, the topologically twisted action of Vafa-Witten theory can be expressed
as a $Q$-exact term $\{Q, I\}$, plus a term multiplying the complexified coupling constant $\tau$:
\begin{equation}
\mathcal{S}_{\text { twisted }}=\{\mathcal{Q}, I\}-2 \pi i \tau(n-\Delta),
\end{equation}
where $n$ denotes the instanton number,
\begin{equation}
n=\frac{1}{8 \pi^{2}} \int_{M} \operatorname{Tr} F \wedge F .
\end{equation}
In fact, the partition function of the topological twisted super Yang-Mills field theory depend only on
$\tau$ and the gauge group $SU(N)$ chosen, which is a modular form \cite{Vafa:1994tf,Labastida:1999ij,Witten:1998wy},
%\begin{equation}
%\begin{array}{l}{Z_{v}(\tau+1)=\mathrm{e}^{-i \pi \frac{N-1}{N} v^{2}} Z_{v}(\tau)} \\ {Z_{v}(-1 / \tau)=N^{-11}\left(\frac{\tau}{i}\right)^{-24} \sum_{u} \mathrm{e}^{\frac{2 i \pi \psi u}{N}} Z_{u}(\tau)}\end{array}
%\begin{array}{l}{Z(\tau+1)=\mathrm{e}^{-i \pi \frac{N-1}{N} v^{2}} Z(\tau)} \\ {Z(-1 / \tau)=N^{-11}\left(\frac{\tau}{i}\right)^{-24} \sum_{u} \mathrm{e}^{\frac{2 i \pi \psi u}{N}} Z(\tau)}\end{array}
%\end{equation}
\begin{equation}
\begin{array}{c}{Z_{S U(N)}(-1 / \tau)=\pm N^{-1+b_{1}-\frac{b_{2}}{2}}\left(\frac{\tau}{i}\right)^{\frac{w}{2}} Z_{S U(N) / \mathbf{Z}_{N}}(\tau)} \\ {=\pm N^{-\chi / 2}\left(\frac{\tau}{i}\right)^{\frac{w}{2}} Z_{S U(N) / \mathbf{Z}_{N}}(\tau)}\end{array}.
\end{equation}
Where $\chi$ is Euler characteristics, $w$ is the modular weight. For any finite value of $N$, the partition function
$Z(\tau)$ is smooth as a function of $\tau$, but for large $N$ limit, as $N \rightarrow \infty$, the function becomes
non-smooth \cite{Maloney:2007ud}. Or, in other words, the function is not a analytic function now. The original idea of
Lee and Yang is that although a system in finite volume can have no phase transition, its partition function, depending
on the complexified thermodynamic variables, can have zeroes. Then, in the infinite volume limit, the zeroes become
more numerous and may become dense. Then, a true phase transition can emerge. In our problem, the limit
$N \rightarrow \infty$, is analogous to a thermodynamic limit. Or, in other words, Lee-Yang type
phase transition happens in the topological twisted super Yang-Mills field theory in the large $N$ limit.

%\section{Summary and conclusion}
In summary, we introduce the Jones polynomial and topological field theory. In fact, the expectation value of the
Wilson loop is the Jones polynomial. Witten find that the Hawking-page phase transition in $Ads_{3}$ space is
Lee-yang type. Then we study the zeros of the topological field theory. And we shown that the zero of hopf link
obey the Lee-Yang unit circle theorem. It is suggested that there is no phase transition in this link.
Since all the Jones polynomials of torus $J\left(m,n\right)$ there are at least two real zeros.
Then we argue that phase transition can happen in the torus knot. Or, more accurately, phase transitions can
happen in the topological field theory which the corresponding knot is torus knot. And we also show that the Lee-Yang type phase transition happens in the topological twisted super Yang-Mills field theory in the large N limit.

The $\frac{1}{N}$ expansion of the free energy of Chern-Simons theory~\cite{Ooguri:1999bv,Neitzke:2004ni}
takes the form $F=\sum_{g, h} C_{g, h} N^{h} \kappa^{2 g-2+h}=\sum_{g, h} C_{g, h} N^{2-2 g} \lambda^{2 g-2+h}$.
However, in the large N limit, it is possible to have phase transitions even in finite volume. Further evidence
was found in that the Wilson lines observable in the topological string theory gives the same knot
invariants~\cite{Labastida:2000yw}. Then there is natural question to ask: Can phase transition really happen
in the topological string theory?

%\section{References}

%Journal: author's name. journal name, year of publication, volume
%number (Issue No.): page number (as shown in Ref.~\cite{lab1})

%Monographs: the author's name. Title. Version (version 1 can be
%abbreviated). Published in: Publisher, publication year. Page No.
%(The format is shown as in Ref.~\cite{lab2})

%Collection: the author's name. Text title. See (in English ):
%Editor. Essays name. Published in: Publisher, Publication year. Page
%No. (The format is shown as in Ref.~\cite{lab3})

%In the text, commands $\backslash $cite\{lab1\} or $\backslash
%$cite\{lab1, lab2, lab3\} is used for citing a single
%reference or a number of references.

%\section{Footnotes}

%Footnotes should be numbered sequentially in superscript
%owercase roman letters.\footnote{Footnotes should be
%typeset in 8~pt  roman at the bottom of the page.}
%\\

\acknowledgments{We thanks for discussing with Xun Chen. Part of the work was done when Jing Zhou visited the Yau Mathematical Science Center.
}

\end{multicols}

\vspace{10mm}

\begin{multicols}{2}

%\subsection*{Appendices A}
%\begin{small}

%\noindent{\bf Subtitle}

%Appendices are generally placed after the references. The equations
%should be numbered as A1, A2, $ \cdots $, and the letter size of the
%text should be 9pt.

%\begin{subequations}
%\renewcommand{\theequation}{A\arabic{equation}}
%\begin{equation}
%\mu(n, t) = {\sum^\infty_{i=1} 1(d_i < t, N(d_i)
%= n)}{\vint\nolimits^{\;t}_{\sigma=0} 1(N(\sigma) = n){\rm d}\sigma}\,. \label{a1}
%\end{equation}

%\begin{equation}
%\mu(n, t) = {\sum^\infty_{i=1} 1(d_i < t,
%N(d_i)=n)}{\vint\nolimits^{\;t}_{\sigma=0} 1(N(\sigma) = n){\rm d}\sigma}\,. \label{a2}
%\end{equation}

%\begin{equation}
%\label{a3}
%F = S \prod \limits_{i < j} \Big \{ \sum_{p=1}^{n} f^{p}(r_{ij})
%O^{p}(i,j) \Big\},
%\end{equation}

%\end{subequations}
%\end{small}

\end{multicols}

\vspace{-1mm}
\centerline{\rule{80mm}{0.1pt}}
\vspace{2mm}

\begin{multicols}{2}

\end{multicols}

\clearpage

%\end{CJK*}
\end{document}